# VERIFICATION TECHNOLOGY FOR FINGER VEIN BIOMETRIC

George Kumi Kyeremeh[1], M. Abdul-Al, R.[1], Qahwaji[1] and R.A. Abd-Alhameed[1],

Faculty of Engineering and Informatics, University of Bradford[1], Bradford, BD7 1DP, UK,

{G.k.kyeremeh@bradford.ac.uk; M.Abdul.Al@bradford.ac.uk; n.a.abduljabbar3@bradford.ac.uk; R.A.A.Abd@bradford.ac.uk; R.S.R.Qahwaji@bradford.ac.uk }

## ABSTRACT

**Finger vein** biometrics is an approach to identifying individuals based on the unique patterns of blood vessels in their fingers. The technology has evolved significantly over the past few decades, with technological advancements in image capture and processing techniques leading to more efficient, accurate, and reliable systems. Earlier methods of finger vein capture involved using infrared cameras to take images of the finger veins, which were compared to a pre-existing database of known individuals. More recent methods include using laser diodes to illuminate the finger veins or using near-infrared light to capture images that are more detailed and accurate.

This article focuses on a verification system that compares the matrices of an efficient finger vein verification system on different databases to test its strength and efficiency. Contrast Limited Adaptive Histogram Equalization (CLAHE) has been examined as an image enhancement and processing method to improve contrast to render details in an image easier to detect, particularly in cases where the image contains uneven lighting. The article further describes the design and testing of an advanced finger vein verification system that uses cutting-edge metrics to measure its effectiveness. To identify and verify people based on their finger vein patterns, the system combines a Random Forest classifier with a comparison between two pretrained systems, VGG16 and ResNet50, which are types of convolutional neural networks. Pretrained weights in Imagenet are deployed, which is a pretrained dataset, making it easier to classify images and reducing the computational complexity of the system. The evaluation measures include the Equal Error Rate (EER), False Accept Rate (FAR), False Reject Rate (FRR), threshold, EER False Match Rate, EER False Non-Match Rate, and confidence. Experimental results demonstrate the system's effectiveness in identifying individuals based on finger vein patterns. VGG16 generally performs better across all databases in this study, with higher confidence levels and accuracy in its predictions compared to ResNet50. While both models perform well on these databases, VGG16 has a slightly lower EER compared to ResNet50, indicating slightly better performance in terms of balancing FAR and FRR. Per our research, this is the first study on a practical verification system with finger vein images. VGG-16 and ResNet-50 models are implemented on three different datasets, and fine-tuning these models enabled the harnessing of their powerful capabilities and achieving superior performance on the specific image classification task.

# 1. INTRODUCTION

Finger vein biometrics (vascular biometrics) is a technology that utilizes the distinctive patterns in the veins of a person's finger to identify them. This technology has been explored for decades and has gained relevance in recent years to secure buildings, devices, and other systems. The first known use of finger vein biometrics dates to the late 1970s [1] when researchers at the Nippon Telegraph and Telephone Corporation (NTT) in Japan developed a system that used infrared light to capture images of the veins in a person's finger. The system was able to accurately identify individuals with a high degree of accuracy, and it was commercialized by NTT in the 1980s as the "Vein Scan" system [2],[3]. Since the discovery, there have been numerous attempts to utilize finger vein biometrics in a variety of applications, including access control, time and attendance tracking, and identification. It can be applied in healthcare settings to identify patients, ensuring that the right treatment is given properly and securely to the right patient, reducing medical errors, and improving patient safety. This may lessen medical mistakes like giving a patient the incorrect drug or treatment based of similarities such as name and appearance. It can be applied to restrict access to places like patient records or spaces for the storage of medications. This can lessen the possibility of errors or misuse by preventing unauthorised access to sensitive information ensuring that only authorized healthcare providers can view or modify records or data. [4], [5]. One major advantage of finger vein biometrics is that it is a contactless technology, meaning that it does not require physical contact with the finger to capture the vein patterns [6]. This makes it hygienic and suitable for use in situations where contact with surfaces or objects may not be desirable.

Over the years, finger vein biometrics technology has undergone significant advancements, resulting in improved accuracy and reliability. These advancements were driven by advances in image capture technology, and machine learning algorithms, among other areas. One key improvement is the use of near-infrared (NIR) imaging technology, which allows for the capture of high-quality images of the veins in a person's finger [7]. NIR imaging works by focusing a NIR light onto the finger, which absorbs the blood in the veins, as shown in Figure 1 and is reflected to the camera. This allows for the capture of detailed images of the vein patterns, which helps to locate/identify individuals with a high degree of accuracy. A further significant development is the use of machine learning algorithms to analyse [7],[8],[2]and compare the images of the vein patterns. These algorithms can learn from a large dataset of images and use that knowledge to accurately find individuals based on their unique vein patterns. Other advancements in finger vein biometrics technology include the use of multi-spectral imaging [9],[10] which captures images of the veins using multiple wavelengths of light, and the development of compact, portable devices that are used to capture images of the veins in a person's finger. The methodology deployed in this study utilizes CLAHE to enhance the finger vein images, compare VGG16 and ResNet50 as feature extractors to extract features from the images, and random forest to classify the images of individuals after they have been captured to build a verification system.

## 1.1 OTHER BIOMETRICS DEPLOYED

, As shown in Table 1 [11]   biometric characteristics can be divided into two classes: feature-based and holistic. The feature-based technique separates low-level components based on their properties and analyses them in terms of spatial placement, whereas the holistic approach recognises the full finger. [12],[13].

*Table 1. Comparison Of Various Biometric Technologies [11]*

| Biometric Characteristics | Universality | Distinctiveness | Permanence | Collectability | Performance | Acceptability | Circumventing |
|---|---|---|---|---|---|---|---|

| Face | High | Low | Medium | High | Low | High | High |
|---|---|---|---|---|---|---|---|
| Facial thermogram | High | High | Low | High | Low | High | Medium |
| Fingerprint | Medium | High | High | Medium | High | Medium | Medium |
| Iris | High | High | High | Medium | High | Low | Low |
| Retina | High | High | Medium | Low | High | Low | Low |
| Hand geometry | Medium | Medium | Medium | High | Medium | Medium | Medium |
| Hand vein | Medium | Medium | Medium | Medium | Medium | Medium | Low |
| Palm print | Medium | High | High | Medium | High | Medium | Medium |
| Voice | Medium | Low | Low | Medium | Low | High | High |
| Keystroke | Low | Low | Low | Medium | Low | Medium | Medium |
| Gait | Medium | Low | Low | High | Low | High | Medium |
| Signature | Low | Low | Low | High | Low | High | High |
| DNA | High | High | High | Low | High | Low | Low |

These characteristics can be defined as [11]:

- Universality: the trait possessed by everyone.
- Distinctiveness: A characteristic should be sufficiently different for each individual.
- Permanence: the attribute must be sufficiently invariant about its usage.
- Collectability: the attribute has a quantifiable measurement.
- Performance: refers to the sum of the possible recognition speed and accuracy, the resources needed to reach the target speed and accuracy, and the operational and environmental variables influencing the speed and accuracy.
- Acceptance: a measure of how much individuals are willing to tolerate a specific biometric characteristic being used on frequent bases
- Circumvention: a measure of how easily fraudulent techniques can be used to mislead the system.

## 2. FINGER VEIN BIOMETRICS

Finger vein biometrics offer numerous advantages over other biometric sense modalities [14],[15],[16]:

1. High accuracy: Finger vein biometrics have high accuracy rates since each person's individual vein patterns are difficult to imitate or forge. Finger vein biometrics are, therefore, more trustworthy, and safer than other biometric modalities.
2. . Finger vein biometrics are now more hygienic and practical for users.
3. Difficult to alter: Finger vein biometric information is kept inside the body, making it challenging for an adversary to take or alter. As other methods store biometric information outside of the body, finger vein biometrics are therefore more secure than those methods.
4. Steady and consistent: Even as a person ages, the vein patterns in their fingers remain stable and consistent over time. Because other biometric modalities can be influenced by alterations in physical appearance or environmental factors, finger vein biometrics are therefore more trustworthy than those modalities.

### 2.1 ANATOMY AND ANALYSIS OF THE FINGER VEIN

A person's finger veins are a component of the circulatory system that controls the return of deoxygenated blood to the heart. They are said to be a network of thin, black lines that are just below the skin. The arteries that carry oxygenated blood from the heart to various regions of the body are joined to the veins in the finger [15]. The tunica intima, tunica media, and tunica adventitia are the three layers that make up the finger's veins, as depicted in Figure 2. The tunica intima, a thin layer of tissue that separates the veins and arteries, aids in preventing the mixing of oxygenated and deoxygenated blood. The finger's veins are tiny, ranging in size from 0.3 to 1.0 mm [15]. The tunica media, a layer of smooth muscle that surrounds the veins, helps control blood flow, and preserve the morphology of the veins as indicated in Figure 1 [17] and Figure 2. [18]. The arrangement and form of the veins, as well as the position of the capillaries, all contribute to the distinctive patterns of the veins in a person's finger [19]. Everyone has a different pattern that can be used for biometric identification [20], [21], [16], [19]. The size, shape, and placement of the veins, together with other elements like the thickness of the skin and the presence of other structures like tendons and bones [19] as seen in Figure 1, combine to form the distinctive patterns in each person's finger veins. Finger vein biometrics are only conceivable because of these patterns, which can be used to distinguishably identify people based on their vein patterns.

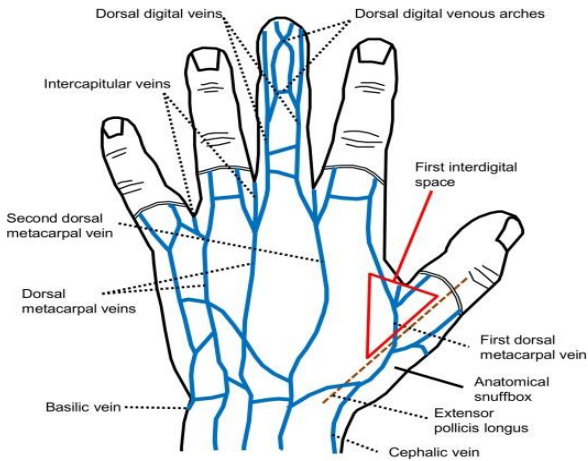 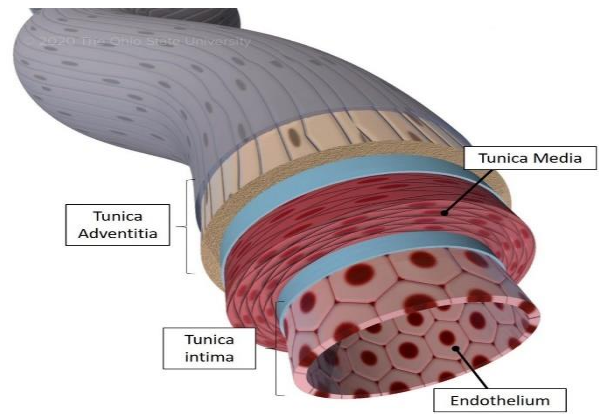

*Figure 1. Vein structure of the hand*                                                                                   *Layers of the finger vein*

### 2.2 Contrast Limited AHE (CLAHE) ENHANCEMENT METHOD

CLAHE is an image processing technique used to enhance the contrast and improve the visibility of details in an image, especially when there are variations in lighting across the image. CLAHE is particularly useful in medical imaging, computer vision, and various other fields where image quality is crucial [22].

CLAHE is based on the concept of **histogram equalization**, [23] which aims to spread out the intensity values in an image's histogram. This helps in utilizing the full dynamic range of pixel values, enhancing the contrast. Unlike traditional histogram equalization, which applies a global transformation to the entire image, CLAHE takes a more **adaptive approach**. It divides the image into small, non-overlapping tiles or blocks. For each of these tiles, CLAHE computes a histogram. This allows the method to adjust the contrast locally, considering the characteristics of each region. **Local histograms** ensure that the contrast enhancement doesn't oversaturate or overamplify certain parts of the image [22]. CLAHE includes a key feature called "**contrast limiting**" [24]. After computing the histogram for each tile, it redistributes the pixel values in a way that limits excessive amplification of the local contrast. This prevents the introduction of unwanted noise and ensures a visually appealing result. To eliminate visible seams between neighbouring tiles, CLAHE employs **interpolation** techniques [25], [23]. This helps in creating a smooth transition between regions with different contrast enhancements.

### 2.3 SUITABILITY OF CLAHE FOR VEINS.

- Enhancement of Local Contrast: Vein images may contain areas with varying levels of contrast around the edges of veins. In such circumstances, CLAHE increases the local contrast, which makes the veins easier to identify and distinguish.
- Adaptive Histogram Equalization (CLAHE): This technique separates the image into small tiles and applies histogram equalization to each tile independently. In vein images, where tissue characteristics or illumination conditions may fluctuate, the adaptive technique guarantees that contrast enhancement is applied accurately to different sections of the image.
- Limiting Over-Enhancement: To avoid over-enhancement, which can result in loss of detail, CLAHE includes a contrast limiting factor. This is especially crucial for vein images since over-enhancement might introduce noise or hide significant features.

- Robustness to Uneven Illumination: Vein images taken in non-uniform lighting may exhibit uneven illumination, which could make certain regions appear darker or brighter than others. CLAHE can help reduce the impacts of uneven illumination and produce a more consistent appearance throughout the image.
- Preservation of Vein Structure: CLAHE contributes to the preservation of the vein patterns' overall structure while augmenting the local contrast. This makes the vein patterns more amenable to later processing stages like segmentation and feature extraction.

## 2.4 METHODS OF FINGER VEIN FEATURE EXTRACTION

Different methods are deployed for finger vein feature extraction with their advantages and disadvantages as described in Table 2. [26], these include the process of locating and extracting the unique features and characteristics of the veins in a person's finger for use in biometric identification.

***Principal component analysis (PCA)*** is a frequently used technique in machine learning, pattern recognition, and other areas where huge datasets are analysed. PCA works by locating a dataset's principal components, or the directions in the data that are most variable. The eigenvectors of the data's covariance matrix are used to calculate these primary components. [27]. The dataset's dimensionality can be decreased by projecting the data onto the principal components once they have been found. This is accomplished by picking the top main components that account for the greatest amount of variation in the data and ignoring the others. [28]. When a dataset has numerous features or variables, making it challenging to analyse or visualise, PCA is especially helpful. PCA can streamline the analysis and display of the data while retaining the most crucial information by reducing the dimensionality of the data. Several domains, including image processing, data reduction, feature extraction, and classification, have applications for PCA. [29]. For example, PCA can be used in the context of finger vein recognition to extract the most crucial features from the images of the vein patterns that have been acquired, decreasing the dimensionality of the data, and enhancing the effectiveness and efficiency of the recognition system.

***Independent component analysis (ICA)***, which entails the division of the finger vein image data into separate components that represent various characteristics of the data, is another approach that is frequently used to extract finger vein features. The shape, size, and placement of the veins, along with other traits that are unique to the individual, can be extracted using ICA. [30],[31]. Independent Component Analysis (ICA) is a statistical method used for separating a multivariate signal into its independent, non-Gaussian components [32]. It is a form of blind source separation that can be used to extract hidden factors or sources that contribute to a given signal. ICA works by assuming that the observed data is a linear combination of several independent sources, each with its distinct probability distribution. The goal of ICA is to estimate the mixing matrix that relates the observed data to the underlying sources, and then use this matrix to separate the sources from the observed data. ICA has a wide range of applications in various fields, including signal processing, image analysis, and machine learning. In the context of finger vein recognition, ICA can be used to extract the most relevant features from the captured images of the vein patterns. This can help improve the accuracy and efficiency of the recognition system, especially in cases where the captured images contain noise or unwanted artifacts [30],[31]. ICA is often used in combination with other techniques, such as PCA or wavelet transforms, to achieve better separation and extraction of the independent components. It is a powerful tool for data analysis and can help reveal underlying patterns and relationships in complex datasets [33].

***Linear discriminant analysis (LDA)*** is a statistical method commonly used for pattern recognition and classification tasks. It aims to find a linear combination of features that maximizes the separation between classes while minimizing the within-class variability. This technique has been successfully applied in various research fields, such as face recognition, spectroscopy, gait pattern classification, and brain tumour classification [34]. In the specific context of finger vein analysis, linear discriminant analysis can be applied to extract relevant features from finger vein images and classify them into different categories based on their unique vein patterns [35]. Linear discriminant analysis works by finding a projection that maximizes the separation between classes in the feature space [36]. This is achieved by calculating the between-class scatter matrix and within-class scatter matrix. The between-class scatter matrix quantifies the separation between different classes by calculating the covariance between their means. The within-class scatter matrix measures the variability within each class by calculating the covariance within each class. The goal of LDA is to find a projection that maximizes the ratio of between-class scatter to within-class scatter, leading to maximum separability between the different classes [37].

*Local binary patterns (LBP).* The Local Binary Patterns feature extraction technique is commonly used to analyse and extract texture patterns from biometric images such as finger vein images [38],[39],[40]. LBP works by dividing an image into small local regions and analysing the pixel intensity values within each region. It then encodes these intensity relationships into binary patterns, which are representative of the texture information present in that region. By considering the spatial relationships between neighbouring pixels, LBP can effectively capture local variations and patterns within an image [41]. In the context of finger vein images, LBP proves to be a valuable tool for feature extraction. Finger veins possess distinct patterns that are unique to everyone, making them suitable for biometric identification. By applying LBP to finger vein images, researchers and practitioners can extract relevant texture features that aid in distinguishing one person's finger vein pattern from another. The extracted LBP features can. Finger vein recognition has gained significant attention in recent years due to its high accuracy and reliability. The distinct patterns found in finger veins offer a robust and secure method for biometric identification [42]. By utilizing Local Binary Patterns (LBP) on finger vein images, researchers and practitioners can extract essential texture features that play a crucial role in distinguishing the unique finger vein patterns of individuals. These extracted LBP features provide a solid foundation for developing efficient and effective finger vein recognition systems.

*Mean Curvature (MC)* is a geometric property which has emerged as an efficient and reliable method used in finger vein feature extraction making it a valuable technique in biometric identification systems. This method entails analysing the finger vein pattern by measuring the average curvature of the vein's structure [43]. MC-based feature extraction focuses on the spatial properties of the vein structure, considering the variations in curvature along the vein patterns. The mean curvature is calculated, which is the average of the curvatures at different points along the vein and aims to capture unique characteristics of the vein patterns, such as the smoothness or irregularity in the vein structure. The MC-based method recognizes and encodes distinctive geometric properties of the vein patterns, which can potentially offer robust and discriminative features for biometric recognition systems [44]. However, the efficiency of MC as a feature extraction method may be influenced by the quality and consistency of the vein images, as well as the ability to handle variations caused by factors such as positioning, image quality, and environmental conditions.

*Table 2. Feature Extraction Methods*

| Ref | Approach | Category | Advantage | Disadvantages |
|---|---|---|---|---|
| [45] | Principal Component Analysis (PCA) | Dimensionality Reduction Based | This method has the advantage of a high recognition rate due to the feature's small dimensions, a reduction that changes the images from a higher to a smaller dimension. | Low recognition rate insufficiency of robustness. Global features are highly dependent on parameters such as location, occlusion, distortion, and lighting and are hence unsuitable for feature extraction. |
| [33] | Independent component analysis (ICA) | | | |
| [46] | Linear discriminant analysis (LDA) | | | |

| [41] | Local binary patterns (LBP). | Local Binary-Based Method | Minimal correlation between characteristics and high robustness. | Time time-consuming during computations and slightly higher dimensions of features. |
|---|---|---|---|---|
| [44] | Mean Curvature (MC) | Vein Pattern Based | Not sensitive to unequal vein width. Improve image quality which enhances recognition. | Incorrectly tuned pixels. Low robustness |

*2.5 LIMITATIONS OF EXISTING METHODS OF FEATURE EXTRACTION METHODS*

The effectiveness of the existing finger vein feature extraction processes does not elude limitations. One significant constraint involves variations in image quality and conditions, as finger vein images are sensitive to factors like illumination changes, finger positioning, and environmental conditions, leading to inconsistencies in captured vein patterns. Additionally, the reliance on specific imaging devices and the need for direct contact during image acquisition limits practical deployment in real-world scenarios [14],. Moreover, the computational complexity of some feature extraction algorithms hinders real-time processing, posing challenges for applications requiring rapid identification [47]. These limitations collectively impact the robustness, generalizability, and real-time applicability of existing finger vein feature extraction methods, highlighting the need for advancements in handling variability, improving robustness, and ensuring adaptability across diverse conditions and demographics.

In addition to the general shortcomings of the existing methods of finger vein processing, specific processes also have peculiar deficiencies as presented in Table. 3 below [26].The feature extraction techniques and matching strategy have an impact on several of the present handcrafted features. Handcrafted features underperform when used with poor-quality images since they are noise-sensitive [48]. In addition, it is challenging to define standard parameters for procedures, and finding the right settings for various databases necessitates extensive experimentation. After feature extraction, the selection of a matching approach can also influence how well a recognition system performs. The deep learning method, which has superseded the handcrafted-based method in the field of computer vision, has been presented for finger-vein recognition to get beyond the traditional handcrafted-based method's limitations. Convolutional neural networks (CNNs) which is a deep learning technique that draws inspiration from the human brain, can be used to learn hierarchical representations and characteristics from data [49],[50]. Deep learning can simultaneously extract features, decrease data dimension, and classify in a network structure, which gives it a substantial edge over conventional methods. Additionally, independent of image quality and finger placement, deep learning techniques can achieve good and reliable recognition performance. For feature extraction, classification, image restoration, and image quality evaluation, deep learning techniques have been employed successfully. The deep-learning-based approach to image quality evaluation has been employed to predict low-quality images. To extract features from binary images for quality assessment and obtain excellent performance in the public database, Qin and El-Yacoubi [51],[52] constructed a CNN based on the presumption that low-quality images are falsely rejected. To determine if an image has complex and stable vein patterns, Zeng et al.[53] used a light-CNN-based image quality assessment method. The image was divided into numerous blocks, and the image quality was then represented by averaging the results from these block images. The suggested light-CNN architecture reduces computation time while simultaneously extracting reliable features to reflect the quality of finger-vein images. The deep-learning-based method can automatically train robust features for evaluating image quality and achieve noticeably improved performance when compared to hand-crafted descriptor-based methods.

Jing et al.[54] used an autoencoder network to restore the tainted image during image restoration. Regarding restoring finger veins, the designed autoencoder network significantly improved the receptive field of convolution by using hole convolution. For the restoration of vein images, the modified conditional generative adversarial network (conditional GAN) was suggested in [55]. Experiments on two publicly accessible databases demonstrate superior performance in image restoration by taking advantage of the robust learning and representation capabilities of GAN. The deep learning-based approach has achieved superior performance in feature extraction and classification thanks to its strong feature learning capabilities. A deep learning model for vein pattern extraction and recovery was created by Qin et al.[56]. Convolutional autoencoder-based techniques to learn the high-level features were introduced by Yan [57], [58]. The proposed method outperformed the conventional dimension reduction method in terms of learning features, according to experimental findings. The training and configuration settings of the autoencoder architectures, U-net, RefineNet, and SegNet, were investigated to ascertain the effectiveness of three various conventional autoencoder designs in extracting finger vein patterns and were assessed by Jalilian and Uhl [59]. The experiments on the SDU database and UTFVP database highlighted the fact that autoencoder architectures with automatically generated labels improved the performance of pattern extraction. Finger vein generative adversarial network (FV-GAN), developed by Yang et al. [60], was a cycle GAN (Cycle-GAN)-based technique for pattern extraction. The FV-GAN was created to directly and intuitively express vein patterns. A CNN network was employed by Song et al. [61] to extract features from the composite image, which was then matched using the shift matching approach. Experiments show that the proposed solution, which uses a composite image and a shift-matching algorithm, is very robust to noise and misalignment.

To extract the vein pattern network for finger-vein verification, Zeng et al. [62] created an end-to-end model that combined the full CNN (FCNN) with a conditional random field (CRF). A straightforward four-layer CNN model was used by Ahmad Radzi et al. [63] for feature extraction. The outstanding performance of this model was demonstrated using the tenfold cross-validation method. The autoencoder networks were used in [64] to depict images of finger veins by making use of feature learning. In the SDU database, this automatic feature extraction and learning technique produced great accuracy. Meng et al. [65] suggested a CNN model based on the conventional AlexNet architecture to extract feature vectors from finger vein images, and the Euclidean distance between two vectors was calculated to assess similarity. The CNN model consists of five convolutional layers and three fully connected layers. Das et al. [48] assessed and confirmed the efficacy of CNN using four public finger-vein databases with varying image quality. The disadvantage of these methods is that numerous experiments are required to determine the CNN structure's architecture. Additionally, a substantial number of images are required for the network training to achieve the requisite performance, which is not feasible. Additionally, because CNN has many trainable parameters, the training period is lengthy and unsuitable for real-time recognition.

The pretrained network has been used in finger-vein recognition to get higher performance. Pretrained models can be used to simplify the training process, but the extracted features may not be of adequate size. A pretrained VGG-16 model was employed by Hong et al. [50] and it was adjusted for recognizing finger veins as deployed in this study. Then, to obtain better performance, certain hybrid feature extraction approaches that combine the handcrafted-based feature extraction methods with the deep-learning-based methods have been presented. However, the redundant parameters as part of these pretrained systems will not be suitable for real-time recognition as in the case of finger vein recognition. All redundant parameters were deactivated in this experiment. The advantages of finger vein recognition are numerous. Not only is it a non-intrusive and hygienic approach, but it is also highly accurate and difficult to spoof. Existing finger vein recognition systems have made significant progress in terms of deep learning-based techniques, presentation attack detection methods, and multimodal-based systems [66].

The ResNet50 model is a deep learning model that is commonly used as a feature extractor. It is a type of convolutional neural network (CNN) that is specifically designed for image recognition tasks. The ResNet50 model has 50 layers, which makes it deeper than most other CNN models. This depth allows the model to learn more complex patterns and features from the input images. The ResNet50 model is based on the Residual Network (ResNet50) architecture, which was introduced in [67]. The ResNet50 architecture introduces skip connections, or shortcuts, between layers in the network. These shortcuts allow the model to learn more complex patterns and features by passing the output of one layer directly to another layer, bypassing some layers in between. The ResNet50 model is pre-trained on a large dataset called ImageNet, which contains over 14 million images from 1000 different classes. This pre-training allows the model to learn general features and patterns from the images in the dataset. The pre-trained ResNet50 model can then be fine-tuned for specific tasks or datasets by training it on the new dataset for a few epochs. In the context of feature extraction, the ResNet50 model can be used to extract features from input images. These features are then used as input to a machine learning model or a deep learning model for a specific task, such as image classification or object detection. The features extracted by the ResNet50 model are high-level and meaningful, which makes it a popular choice for feature extraction tasks.

However, there is still room for improvement in the field. One area that can benefit from enhancement is the image acquisition and enhancement stage. Current finger vein recognition systems often struggle with poor image quality due to variations in lighting conditions and surface conditions [68]. To address this issue, an approach involving contrast-limited Adaptive Histogram Equalization can be implemented. CLAHE equalization is a technique that enhances the visibility of vein patterns by enhancing the contrast in low-light conditions [23]. By incorporating CLAHE equalization into the finger vein recognition system, the quality of the captured finger vein images can be significantly improved. Once the finger vein images have been enhanced, the next step is to extract robust features from them. This can be achieved by utilizing the VGG16 model, a powerful deep-learning architecture known for its ability to extract high-level features from images compared with ResNet50 which is a powerful CNN model. Both models can be used to extract discriminative features from the pre-processed finger vein images, allowing for more accurate classification. To further improve the accuracy of the finger vein recognition system, the extracted features can be fed into a random forest classifier. The random forest classifier is a versatile and powerful machine-learning algorithm that can effectively handle complex and high-dimensional data [69]. By utilizing the random forest classifier, the system can classify the finger vein patterns with high accuracy and robustness. The integration of CLAHE equalization, VGG16 or ResNet50 for feature extraction, and random forest for classification offers a superior methodology for finger vein extraction compared to existing methods. With the rapidly advancing field of finger vein recognition, there is an increasing need for improved image acquisition and enhancement techniques. The combination of contrast-limited adaptive histogram equalization, with either VGG16 or ResNet50 for feature extraction, and random forest for classification can offer a promising solution.

*Table 3. Shortcomings of the existing feature extraction systems.*

| Ref. | Method | Disadvantages | Matching |
|---|---|---|---|
| [70] | Collaborative feature extraction and cross-correlation matching using Personalized Best Bit Maps (PBBM) | A necessity in the reduction in the energy efficiency threshold value and a reduction in the value of the threshold value. | Conventional recognition |
| [71] | Morphological expansion and filtering with thresholding of entropy for and matching of the template | Misalignment and finger vein darkening influence recognition performance | Conventional recognition |
| [72] | Gray and size normalization with ROI extraction, Customized Local Line Binary Pattern (CLLBP) and matching score | Basically, deals with the acquisition direction improvement of the image | Conventional recognition |
| [73] | Local Binary Pattern (LBP) and SVM for matching | Decreased information to eliminate point characteristics such as bifurcation and the ends of finger vein lines, and the need to expand the data set to include ages and genders | Machine Learning |
| [74] | PCA and Linear Discriminant Analysis (LDA), as well as matching through SVM and an Adaptive Neuro-Fuzzy Inference System (ANFIS) | Works mainly in a controlled environment such as background noise and damage due to poor lighting, observation angle, and other parameters | Machine Learning |
| [75] | A variance of local binary patterns (LBPV), Gaussian filter, and Global Matching SVM | To compute the SVM, each input data set requires its technique of feature extraction and dimension reduction | Machine Learning |
| [76] | Information capacity, a gradient in the spatial domain, entropy with Image contrast and Gabor feature and matching by Support Vector Regression (SVR) | Focus on integrating and creating an ROI in the venous system. A multimodal scheme as many biometric systems | Machine Learning |
| [77] | Location on a grid, feature-level fusion by Fractional Firefly (FFF) and matching by K-Means Support Vector Regression (K-SVM) | Various objective functions are required to be developed to find the ideal weight score and to improve results | Machine Learning |
| [78] | Discriminant Analysis (DA) and k-NN | Low robustness and not using all data set | Machine Learning |
| [31] | Adaptive K-nearest Centroid Neighbor (Ak-NCN) | Not using data set for two sessions, but one session from data | Machine Learning |
| [56] | Fully Convolutional Network (FCN) | Much more remains to be done to improve verification accuracy | Deep Learning |
| [56] | Extraction and patching of Gabor features - Deep Neural Network (DNN)+ Probabilistic Support Vector Machines (P-SVM) | Not to use representative learning in other stages of biometrics of finger veins, i.e., background vein segmentation and verification | Deep Learning |
| [52] | Five convolutional layers, three max-pooling layers, one SoftMax layer, and one ReLu layer with contrast-limited adaptive histogram equalization make up | Cannot be used on images of non-trained classes' finger veins. | Deep Learning |

| | the Convolutional Neural Network (CNN) Model (CLAH) | | |
|---|---|---|---|
| [48] | 4 CNN model | Data augmentation may be used to enhance training samples for four datasets using non-publicly available data to reduce over-customization of the CNN designs | |
| [79] | CNN and k-NN | Results need to be improved because grading takes a long time | |
| [80] | Three convolutional layers, three max-pooling layers, and two fully linked layers comprise the CNN model (FCL) | The suggested system is not robust, and you should improve the performance accuracy. Second, the details of this model should be enhanced and supplemented with the loss function in trials to enable comparison of comparable performance and study of the benefits and drawbacks of each loss function. | |

## 3. PROPOSED FINGER VEIN SYSTEM

The existing finger vein recognition systems have made significant progress in terms of deep learning-based techniques, presentation attack detection methods, and multimodal-based systems. However, there is still room for improvement in the field. One area that can benefit from enhancement is the image acquisition and enhancement stage as reported in [73], [74]. Incorporating transfer learning from pre-trained VGG16 or ResNet50 can further enhance the performance of finger vein biometrics to deal with the issue of robustness as reported by [78] and [80]. VGG16 and ResNet50 are deep convolutional neural networks which are known for their abilities to extract high-level features from images [81]. By leveraging its pre-trained weights, we can effectively capture intricate patterns and unique characteristics present in finger vein images. This not only improves accuracy as it was a shortcoming in [56] but also enables the system to generalize well across different individuals. Further experiments were performed with Random Forest (RF) as a classifier.

### 3.1 Dataset Description

All the databases Table . were split into a training/Learning and testing set. The data used for the training depicted in Table 5 is different from the data used in testing They are enhanced by first reading the image in grayscale format and adjusting the contrast and brightness of the image. The images are further enhanced using the Contrast Limited Adaptive Histogram Equalization (CLAHE) method and a Gaussian filter is applied to the final image.

*Table 4. Details of the database for the study*

| Database | Subjects | No. of Fingers | Details of fingers | Image per finger | Image Size | Total Images |
|---|---|---|---|---|---|---|
| FV-USM | 123 | 4 | L&R (Index & middle) | 6 | 640*480 | 2952 |
| UTFVP | 60 | 6 | L&R (Middle, Index &Ring) | 6 | 672*380 | 1440 |
| PLUSVein-FV3 | 60 | 6 | L&R (Middle, Index &Ring) | 6 | 736*192 | 1800 |

**FV-USM**

This dataset contains 2952 finger vein images, having a resolution of 640*480 pixels obtained from 123 subjects. Two sessions were conducted from the same people for the collection of this database which was separated by more than two weeks' time but data from only one session is used in this study. Four (4) fingers with six (6) images from each finger were captured by a custom-built capturing device consisting of 850nm LEDs, a NIR pass filter and a Sony PSEye camera [82].

**UTFVP**

This database contains 1440 Palmer vein images, and it was captured with the light transmission method. The vein images have a resolution of 672*380 pixels obtained from 60 subjects. Six (6) fingers with four (4) images from each finger were captured by a custom-built capturing device with 850nm LEDs, a 930nm NIR pass filter and a C-Cam BCi5 monochrome camera [83].

**PLUSVein-FV3**

This database contains 1800 Palmer finger vein images with a resolution of 736*192 for the region of interest (ROI) images from 60 subjects. Six (6) fingers with five (5) images from each finger were captured with a capturing device based on an IDS UI-ML1240-NIR camera with an 850nm LEDs and an NIR pass filter [84].

The databases for this study were split into two. A training set (which was used for learing) and a testing set, and the data sample that were used for the learning are different from the testing set.

*Table 5. Visual Display of the Databases used.*

| DATABASE | VGG | RESNET50 |
|---|---|---|

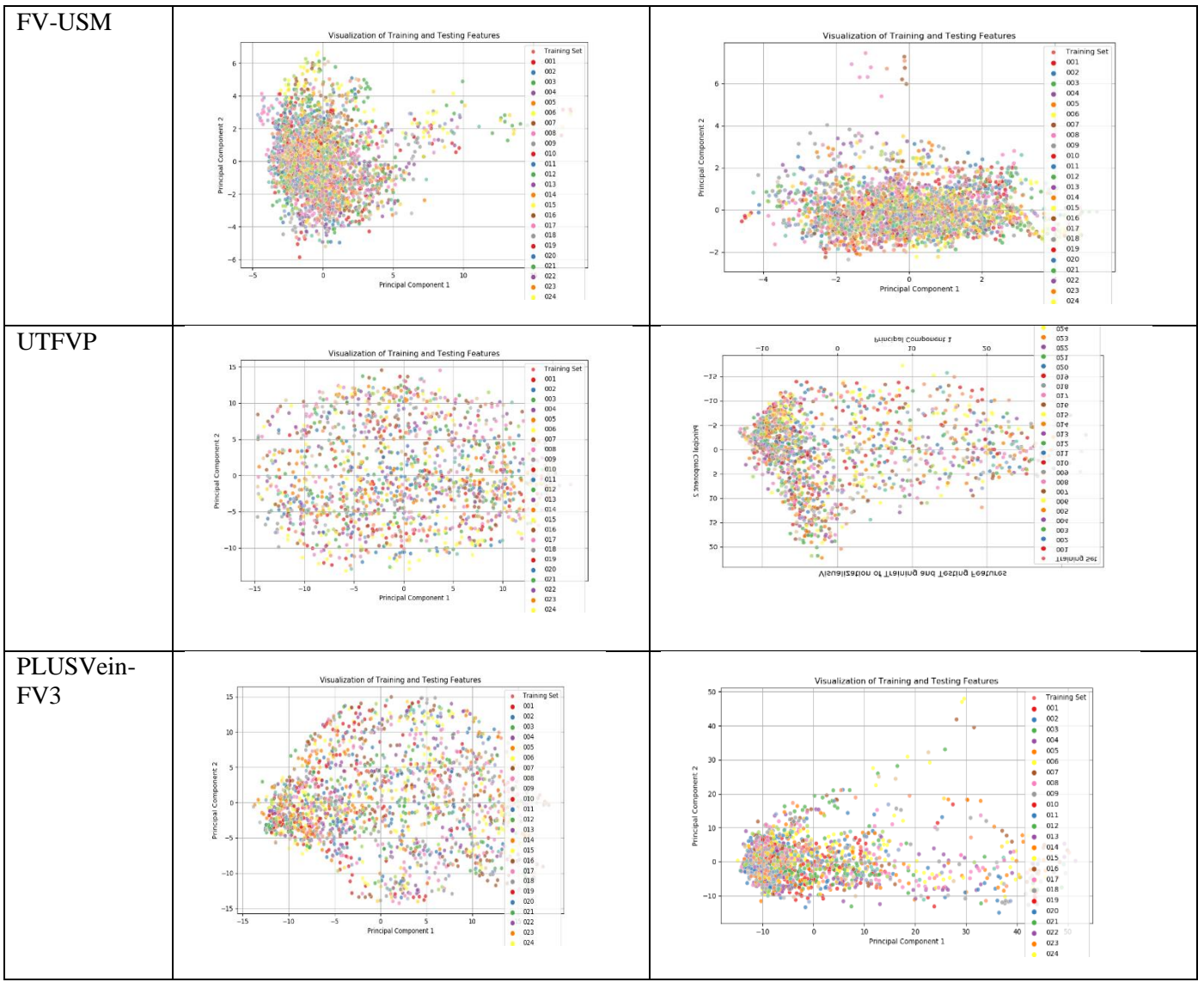

### 3.2 VGG16

Oxford University introduced one of the most popular deep learning architectures which is VGG16 [85]. Due to the comprehensive training, the VGG16 network provides exceptional accuracy even with small image data sets. Transfer learning is been applied here where the pretrained VGG16 has already learned how to detect generic features and patters, whether they're edges or roundedness or whatever those generic features are, it already knows how to detect them and that can be a leverage [86] for image recognition in this study.

As with any pre-trained models, VGG16 [86] requires heavy training if the weights are initialized randomly. In general, Convolutional Neural Network (CCN) models utilize transfer learning (TL) techniques. TL refers to a mechanism in which a model trained on one task is utilized on another task andin this study the weights form *ImageNet* is utilized.

The VGG Net-16 is composed of 16 convolutional layers, 5 pooling layers, and 3 FCLs. In the 1st convolutional layer, 64 filters of size 3×3 are used. Hence, the size of the feature map is ***256×256×3*** for this study in the 1st convolutional layer, where 256 and 256 are the height and width of the feature map, respectively. The photographs are resized to dimensions of 256 by 256 in order to be used as input for the model. The pixel values are normalised since they range from 0 to 255. The weights pertaining to ImageNet are obtained and encompass the top values, which are set to false. Consequently, the dense layers are limited to the final pooling layer, indicating that there is no training process as it progresses through successive epochs with constant weights.

### 3.3 RESNET50

The pre-trained deep learning models, Residual Neural Network (ResNet-50 model is a deep learning model that is commonly used for computer vision. It is a type of convolutional neural network (CNN) that is specifically designed for image recognition tasks. The ResNet50 model has 50 layers, which makes it deeper than most other CNN models. The

ResNet50 model is based on the Residual Network (ResNet50) architecture, which was introduced by [67]. The ResNet50 architecture introduces skip connections, or shortcuts, between layers in the network.

These shortcuts allow the model to learn more complex patterns and features by passing the output of one layer directly to another layer, bypassing some layers in between. The ResNet50 model is pre-trained on a large dataset called ImageNet, which contains over 14 million images from 1000 different classes.

### 3.4 Model Structure

The widely-used VGG-16 and ResNet-50 deep learning models are fine tuned and the newly derived model will learn from the dataset provided , by downloading the weights from Imagenet, and then combined with an output layer, which is a random forest instead of a dense layer in this study. The goal of fine-tuning is to apply the acquired representations from these pre-trained models and modify them according to the finger vein dataset that is pertinent to identification. By leveraging the information captured by these models on the general finger vein image features, better performance and faster convergence can be achieved. Not all of the features of the pre-trained models are utilized during as this part of the fine-tuning procedure. It also help to retain the learned representations of some layers, which are more generic and transferable while allowing the later layers to be adapted to the specific dataset for both VGG-16 and ResNet-50 models. The aim of fine-tuning these models is to gain improved performance on the particular image classification assignment by utilizing their potent capabilities.

### 3.5 RANDOM FOREST AS CLASSIFIER

Since we are comparing VGG16 and ResNet50 in this study Random Forest (RF) is deployed as the classifier. Random Forest has a better performance in comparison to a powerful learner like the Support Vector Machine in several classification problems [87] and it is defined as the ensemble learning system made up of several decision trees [88]. This classifier works by resampling randomly the training set with replacement. The purpose is to generate a new training set which uses a bootstrap resampling method. Each tree which composes Random Forest, consequently, consists of a different subset and is followed by selecting the terminal nodes. For the last process, the result of classification is produced from a majority vote. The fundamental idea of Random Forest is the Gini index minimum, which is described in the following equation,

$$Gini(s) = \sum_{i=1}^{N} q_{ni}(1 - q_{ni})$$

*(eqn) 1*

Where $q_{ni} = \frac{Z_{ni}}{Z}$

where N and $q_{ni}$ are the classes number and the probability of the data related to class *ni* at node t. $z_{ni}$ refers to the tree number for class *ni*.

Theoutput of the learning model is used as an input to the RF model for prediction.. From this level metrics such as accuracy can be determined by comparing the test labels with the predicted by the random forest. Classification in deep learning for biometric image processing is important because it is the basic process of identifying people based on biometric features. This model is state-of-the-art and can play an important role in personal identification and security by accurately identifying biometric data like the finger vein in this study.

### 3.6 VERIFICATION PROCESS

During the verification phase, pre-trained VGG16 and ResNet50 models are used separately to learn from the image from the test dataset that has been enlarged to add a batch dimension. After that, the features are flattened and fed into a Random Forest model that was trained using patterns of finger veins. The trained Random Forest model is used to generate the prediction, and the original label is acquired by performing the inverse label transformation. The anticipated image and the real person who matches the chosen test image are printed together. A random image from the test dataset bearing the anticipated label is then shown, enabling visual confirmation of the prediction The systems performed at different thresholds for the various datasets as they helped to classify matching scores into genuine or imposter categories. In terms of mathematical operations, features are extracted using VGG and ResNet50 models for feature extraction and the final prediction is made using a Random Forest model. Indexing, reshaping, and prediction are executed by machine learning models.

4. EXPERIMENTAL RESULTS AND DISCUSSION

All functions are implemented with Python on PyCharm (IDE) python interpreter on the Lenovo 11[th] Gen Intel CPU and an Intel Core i71165G7 - 2.8 GHz processor. This study considers the matrices that would be useful for a biometric verification system. The results discussed range from the ROC curve and the AUC, the confidence matrix, Equal error rates, false match, and non-match rates.

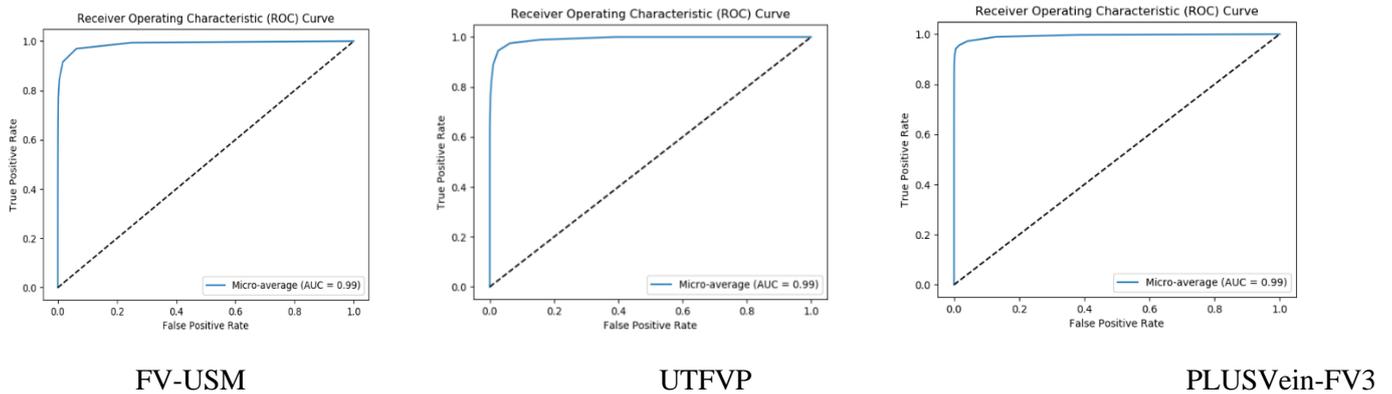

FV-USM    UTFVP    PLUSVein-FV3

*Figure 2. The ROC curve of the corresponding training-validation set in the VGGNet16 model*

.

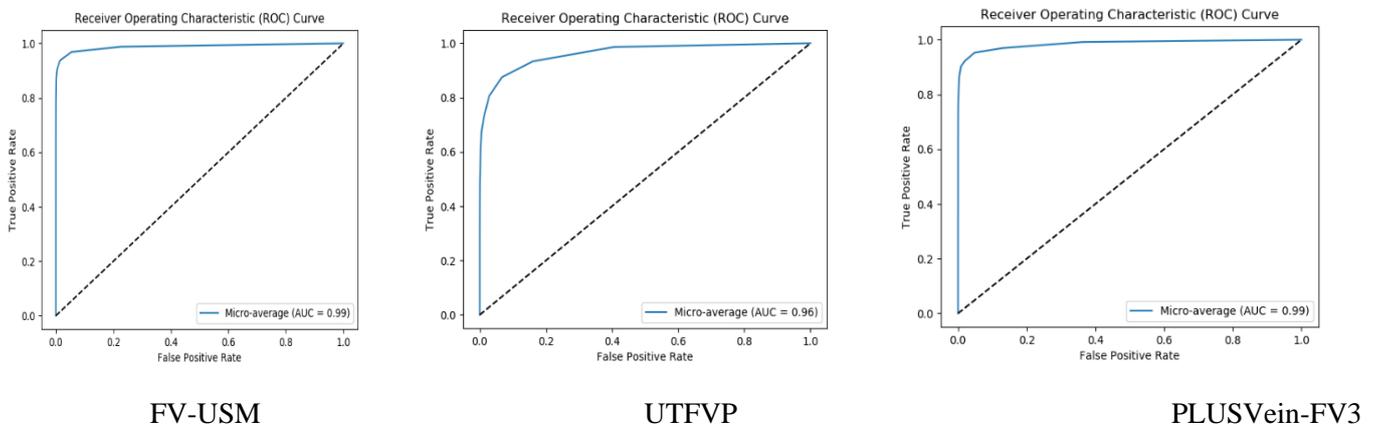

FV-USM    UTFVP    PLUSVein-FV3

*Figure 3. The ROC curve of the corresponding training-validation set in the ResNet50 model*

The system used in tbe study attempts to predict a test label based on the learning done with the training labels. Matrices such as True Positive (TP), True Negative (TN), False positive (FP) and False negative (FN) are determined based on the algorithm shown in eqns 2-5.

$TP = sum((prediction == test\ label)\&(prediction == "genuine"))$

*(eqn) 2*

$TN = sum((prediction == test\ label)\&(prediction != "genuine"))$

*(eqn) 3*

$$FP = sum\big((prediction != test\ label)\&(prediction == "genuine")\big)$$

*(eqn) 4*

$$FN = sum\big((prediction != test\ label)\&(prediction != "genuine")\big)$$

*(eqn) 5*

Where:

TP: Counts the number of instances where both the prediction and the test label are "genuine".

TN: Counts the number of instances where both the prediction and the test label are not "genuine".

FP: Counts the number of instances where the prediction is "genuine" but the test label is not.

FN: Counts the number of instances where the prediction is not "genuine" but the test label is.

True positive rates and False positive rates are further calculated using the the equations 6 and 7 respectively.

True Positive Rate (Sensitivity):

$$TPR = \frac{TP}{TP + FN}$$

*(eqn) 6*

False Positive Rate (1 - Specificity):

$$FPR = \frac{FP}{FP + TN}$$

*(eqn) 7*

The area under the curve (AUC) is further calsulated as:

$$AUC = \sum_{i-1}^{n} \frac{[TPR(i) + TPR(I-1)] * [FPR(i) - FPR(i-1)]}{2}$$

*(eqn) 8*

**Where**

**n**: is the number of thresholds (number of points on the ROC curve).

**TPR[i]**: True Positive Rate at threshold i.

**FPR[i]**: False Positive Rate at threshold i.

**TPR[i-1]**: True Positive Rate at threshold i-1.

**FPR[i-1]**: False Positive Rate at threshold i-1.

This formula eqn 8 determines the Area Under the Curve (AUC) using the trapezoidal rule based on True Positive Rate (TPR) and False Positive Rate (FPR). The AUC is calculated by summing up the areas of trapezoids formed by adjacent points on the ROC curve. Each trapezoid's area is the product of the average height (average of TPRs at adjacent thresholds) and the width (difference in FPRs at adjacent thresholds).

*Table 4. Thresholds, FMR and FNMR for VGG16*

| Dataset | Threshold | False Match Rate (FMR) | False Non-Match Rate (FNMR) |
|---|---|---|---|
| FV-USM Dataset | 0.04 | 0.01 | 0 |
| UTFVP Dataset | 0.16 | 0 | 0 |
| PLUSVein-FV3 Dataset | 0.04 | 0.07 | 0.17 |

*Table 5. Thresholds, FMR and FNMR for ResNet50*

| Dataset | Threshold | False Match Rate (FMR) | False Non-Match Rate (FNMR) |
|---|---|---|---|
| FV-USM Dataset | 0.02 | 0.06 | 0.12 |
| UTFVP Dataset | 0.12 | 0 | 0 |
| PLUSVein-FV3 Dataset | 0.02 | 0.06 | 0.12 |

$$FMR = \frac{FP}{FP + TN}$$

*(eqn) 9*

$$FNMR = \frac{FN}{FN + TP}$$

*(eqn) 10*

FNMR (eqn 9) can also be expressed as False Rejection Rate (FRR) and it is complementary with TPR is the True Positive Rate (or True Match Rate).

$$FRR + TPR = 1$$

*(eqn) 11*

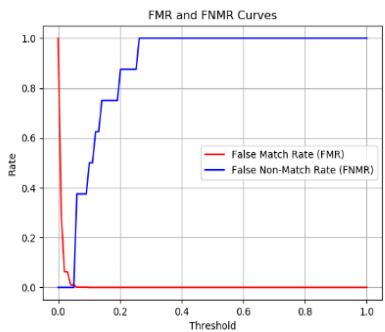  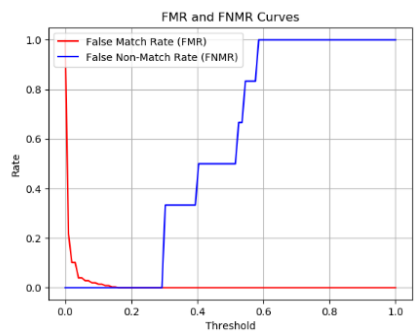  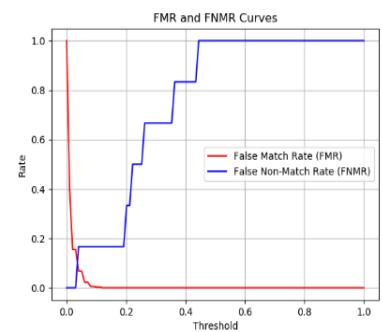

| FV-USM | UTFVP | PLUSVein-FV3 |

*Figure 4. Metrics comparison for the databases with VGG16*

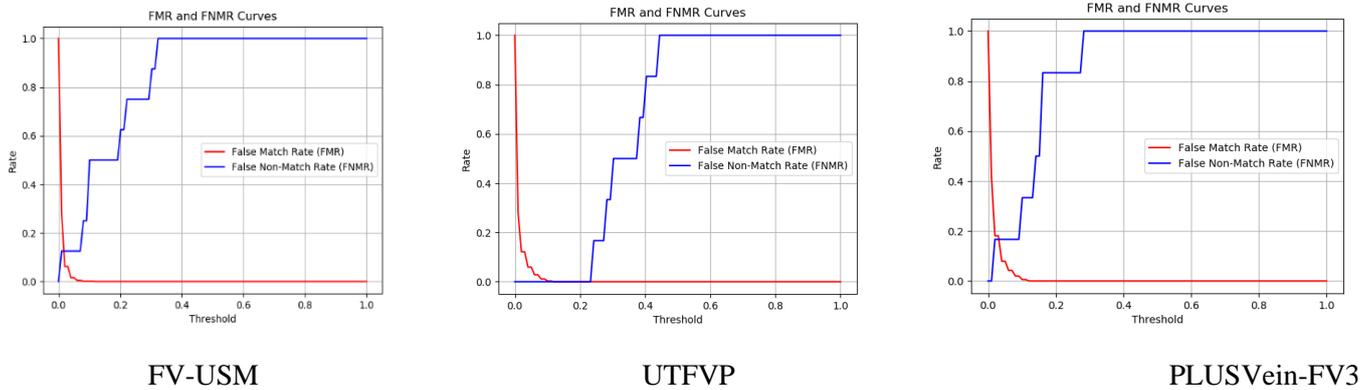

| FV-USM | UTFVP | PLUSVein-FV3 |

Figure 5. Metrics comparison for the databases with ResNet50

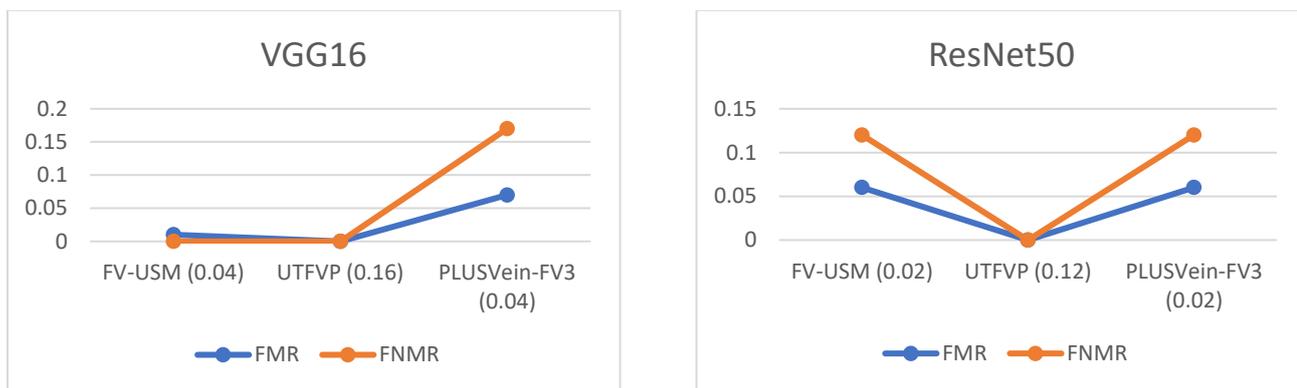

Figure 6. Comparison of the various thresholds

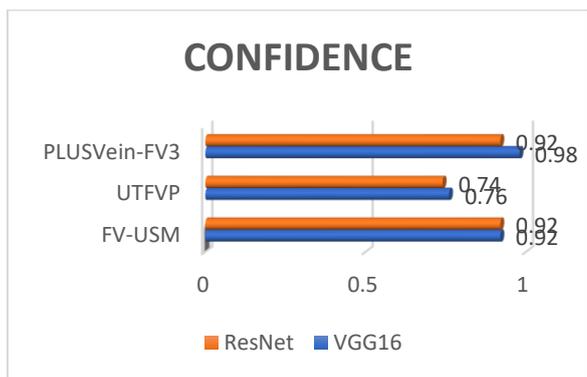

Figure 7. Comparison of the various Confidence matrices

Table 6. Confidence Matrices for VGG16 & ResNet50

| DATABASE | VGG16 | ResNet50 |
|---|---|---|
| FV-USM | 0.92 | 0.92 |
| UTFVP | 0.76 | 0.74 |

| PLUSVein-FV3 | 0.98 | 0.92 |

$$EER = \frac{FAR + FRR}{2}$$

*(eqn) 12*

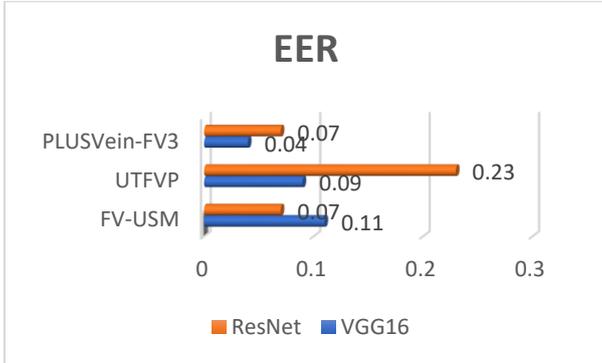

*Figure 8. Comparison of the various EER*

*Table 7. EER for VGG16 & ResNet50 50*

| DATABASE | VGG16 | ResNet50 |
|---|---|---|
| FV-USM | 0.11 | 0.07 |
| UTFVP | 0.09 | 0.23 |
| PLUSVein-FV3 | 0.04 | 0.07 |

### 4.1 RECEIVER OPERATOR CURVE AND AREA UNDER THE CURVE

The micro average AUC (Area Under the Curve) is a performance metric used to evaluate the effectiveness of a binary classifier. It measures the probability that a classifier will rank a positive example higher than a negative example. In this study, the micro average AUC is calculated for each of the three feature extraction methods on the databases FV-USM, UTFVP, and PLUSVein-FV3. The micro average AUC is calculated by averaging the AUC values for each class in the dataset.

FV-USM and PLUSVein-FV3: For these databases, VGG16 and ResNet50 achieved a very high micro average AUC 0.99. This indicates that both models perform exceptionally well on these specific databases, demonstrating high accuracy and discriminative power in classifying samples from different classes. UTFVP: VGG16 achieves a micro-average AUC of 0.99 on the UTFVP database, while ResNet50 achieves a slightly lower micro-average AUC of 0.96. This suggests that VGG16 performs slightly better overall in terms of classification accuracy and discriminative ability on the UTFVP database compared to ResNet50. The micro-average AUC results suggest that VGG16 generally performs slightly better than ResNet50 across the databases, particularly on the UTFVP database where there is a noticeable difference in performance. However, both models exhibit exceptional performance overall, with very high micro-average AUC values across all databases.

Figures 4 & 5 show well distributed with all three databases in a well-distributed ROC curve, where the true positive rate increases rapidly with a low increase in the false positive rate, indicating that the model can effectively discriminate between positive and negative cases. This suggests that the learning done on the features extracted from the finger vein images are informative and allow the model to distinguish between different classes accurately. Compared to the work done by [89], our models have less computational complexities and use a pretrained database "ImageNet" designed which is used in visual object recognition and makes this model more suitable for a verification system.

### 4.2 FALSE MATCH RATE AND FALSE NON-MATCH RATE FOR VGG16

Thresholds are fundamental in biometric verification systems as they strike the balance between false acceptance and false rejection rates, thereby affecting the overall performance and overall security of the system in this study. The systems performed at different thresholds for the various datasets as in Table 6. helped to classify matching scores into genuine or imposter categories. The system is set to determine an a optimal threshold that would operate between the maxima and the minima. This would ensure that the system is not too rigid that would make verification cumbersome or too lenient to breach security.

In the FV-USM Dataset, a threshold of 0.04 was set by the system which means that any matching score above this threshold is classified as genuine, while scores below are classified as imposters. The low FMR of 0.01 indicates that only 1% of imposters are incorrectly accepted as genuine rendering the system and algorithm very secure for this batch of dataset. There are no false rejections ensuring genuine attempts are correctly accepted.

In UTFVP Dataset a threshold of 0.16 was set which is higher than in the FV-USM dataset, indicating a stricter classification criterion. With an FMR and FNMR of 0, this threshold achieves perfect classification, where no imposters are incorrectly accepted, and no genuine attempts are falsely rejected demonstrating very great accuracy and security.

PLUSVein-FV3 Dataset: Like the FV-USM dataset, a threshold of 0.04 is used, but with different error rates. The FMR of 0.07 indicates that 7% of imposters are incorrectly accepted as genuine, which is higher than the FV-USM dataset, suggesting lower security. The FNMR of 0.17 indicates that 17% of genuine attempts are incorrectly rejected, which is also higher compared to the other datasets, representing a lower accuracy.

### 4.3 FALSE MATCH RATE AND FALSE NON-MATCH RATE FOR ResNet50

For FV-USM and PLUSVein-FV3 Dataset with the ResNet50, a threshold of 0.02 will only classify a sample as genuine if its matching score exceeds this threshold. The FMR of 0.06 indicates that 6% of impostor attempts are incorrectly accepted as genuine. The FNMR of 0.12 implies that 12% of genuine attempts are incorrectly rejected. In practical terms, this threshold strikes a balance between security (low FMR) and usability (acceptable FNMR). The FMR and FNMR values are also identical to PLUSVein-FV3, suggesting comparable performance in terms of security and usability.

UTFVP: A threshold of 0.12 means that the system is more permissive in accepting matches compared to FV-USM. It will only classify a sample as genuine if its matching score exceeds this higher threshold. The absence of FMR and FNMR indicates that, at this threshold, the system achieves perfect classification, with no impostor attempts incorrectly accepted as genuine and no genuine attempts incorrectly rejected. This threshold offers high security but may come with usability trade-offs if it's too stringent.

From the threshold in Tables 7 and Figures 6. discussed for the various datasets, if the image resolution is lower (UTFVP), the deep learning model will learn more deep features from the image; on the other hand, if the image resolution is higher, (FV-USM, PLUSVein-FV3) the deep learning model will learn fewer deep features from the image. In a biometric verification system, the choice of threshold is a trade-off running between security and convenience. Relatively lower thresholds improve security by reducing false acceptances but may increase false rejection rates, leading to inconvenience for users. Higher thresholds on the other hand reduce false rejections but may increase the risk of false acceptances. Understanding and optimizing thresholds are essential for achieving the desired balance between security and usability in biometric systems.

### 4.4 CONFIDENCE MATRIX

The confidence matrix results provided looked at how well two different models, VGG16 and ResNet50, did on three different databases: FV-USM, UTFVP, and PLUSVein-FV3 as shown in Figure 7 and Table 8. The values in the matrix represent the confidence level or accuracy of the models in making predictions for each specific database. FV-USM Database VGG16: 0.92, ResNet50: 0.92. Both VGG16 and ResNet50 achieve a high confidence level of 0.92 on the FV-USM database. This indicates that both models perform very well on this database, demonstrating a high accuracy in their predictions. For UTFVP Database VGG16: 0.76, ResNet50: 0.74. VGG16 achieves a confidence level of 0.76 on the UTFVP database, while ResNet50 achieves a slightly lower confidence level of 0.74. This recommends that VGG16 performs slightly better than ResNet50 on this database, with a higher accuracy in its predictions. Then PLUSVein-FV3 Database. VGG16: 0.98, ResNet50: 0.92. VGG16 achieves an exceptionally high confidence level of 0.98 on the

PLUSVein-FV3 database, indicating very accurate predictions. Alternatively, ResNet50 achieves a slightly lower confidence level of 0.92. This implies that VGG16 significantly outperforms ResNet50 on this specific database as well, demonstrating a higher accuracy and confidence in its predictions. The confidence matrix results show the comparative performance of VGG16 and ResNet50 on different databases. VGG16 generally performs better across all databases, with higher confidence levels and accuracy in its predictions compared to ResNet50. However, the extent of the performance difference varies depending on the specific database.

*4.5 EQUAL ERROR RATE*

The Equal Error Rate (EER) figure 8 table 9 is a metric usually used to estimate the performance of biometric authentication and recognition systems. It represents the point on the Receiver Operating Characteristic (ROC) curve where the false acceptance rate (FAR) equals the false rejection rate (FRR). A lower EER indicates better performance, as it signifies a smaller gap between the FAR and FRR. FV-USM Database: VGG16: 0.11, ResNet50: 0.07. In the FV-USM database, the EER for VGG16 is 0.11, while for ResNet50, it is 0.07. This suggests that the ResNet50 model performs better on this database, as it achieves a lower EER compared to VGG16. A lower EER indicates that ResNet50 has a smaller gap between the false acceptance rate (FAR) and false rejection rate (FRR), resulting in more accurate and balanced predictions. UTFVP Database: VGG16: 0.09, ResNet50: 0.23. For the UTFVP database, VGG16 achieves an EER of 0.09, while ResNet50 has a significantly higher EER of 0.23. This indicates that VGG16 outperforms ResNet50 on this database, as it achieves a lower EER and thus demonstrates a better balance between FAR and FRR. PLUSVein-FV3 Database: VGG16: 0.04, ResNet50: 0.07. In the PLUSVein-FV3 database, both VGG16 and ResNet50 achieve relatively low EER values of 0.04 and 0.07, respectively. While both models perform well on this database, VGG16 has a slightly lower EER compared to ResNet50, indicating slightly better performance in terms of balancing FAR and FRR.

The comparison of EER values across different databases indicates variations in the performance of VGG16 and ResNet50. While ResNet50 performs better on the FV-USM database, VGG16 outperforms ResNet50 on the UTFVP database. However, both models generally demonstrate good performance with relatively low EER values on the PLUSVein-FV3 database. The studies by *[90]* described the UTFVP finger vein dataset as inferior compared to the other datasets but this methodology has been able to successfully extract features for recognition with low EER and high throughput even though it was still found to have the lowest confidence among the databases for our system.

*5. CONCLUSION*

This finger vein research is the first study on a practicalize verification system on finger vein images per our research. This verification system performs well and strikes the ideal balance between acceptance and rejection rates due to the use of advanced metrics. The system's accuracy and dependability in recognizing people based on their distinctive finger vein patterns are highlighted by the obtained EER, FAR, FRR, and confidence metrics. It can maintain a strict verification procedure while minimizing false matches, as evidenced by the low EER threshold and False Match Rate. In real-world applications where safe and reliable biometric verification is crucial, the system's high levels of precision and confidence are factors in its effectiveness. The proposed method uses CLAHE to enhance the finger vein images and compared VGG16 and ResNet50 fine-tunned learning models on images, and random forest to classify them. The proposed approach also demonstrates its robustness and applicability in real-world scenarios Capitalizing on less computational complexities and transfer learning, this model is more suitable for a verification system. Relatively lower thresholds improve the security of any biometric verification system by reducing false acceptances but may increase false rejection rates, leading to inconvenience users with several/multiple enrolments in a bid to tighten security. Higher thresholds on the other hand reduce false rejections but may increase the risk of false acceptances. Understanding and optimizing thresholds are essential for achieving the desired balance between security and usability in biometric systems. VGG16 generally performs better across all databases in this study, with higher confidence levels and accuracy in its predictions compared to ResNet50. While both models perform well on these databases, VGG16 has a slightly lower EER compared to ResNet50, indicating slightly better performance in terms of balancing FAR and FRR. The studies by Kauba described the UTFVP finger vein dataset as being inferior compared to the other datasets but this methodology has been able to successfully extract features for recognition with low EER and high throughput even though it was still found to have the lowest confidence of the system. In conclusion, this study addresses the field of finger vein biometrics verification by the application of CLAHE which has facilitated the improvement in the quality of the finger vein images and developed a more efficient and scalable system for finger vein recognition that can manage large-scale deployments and cater for various user groups, with a better Equal Error Rate and still maintain high accuracy and a lower computational time to make it suitable for an effective and biometric functions.

Acknowledgments

This project has received funding from the European Union's Horizon-MSCA-RISE-2019-2023, Marie Skłodowska-Curie, Research, and Innovation Staff Exchange (RISE), titled: Secure and Wireless Multimodal Biometric Scanning Device for Passenger Verification Targeting Land and Sea Border Control.

I also wish to acknowledge the funds from the Government of Ghana through the Ghana Scholarship Secretariat (GSS) in supporting my study at Bradford University.

In addition, the author would like to thank Professor Raymond Veldhuis from the University of Twente, The Netherlands, for his support in using UTFVP database in this work.